\documentclass{elsarticle}
\usepackage[super,compress]{cite}
\usepackage{amsmath,multirow}
\journal{Int. J. of Mod. Phys. E, DOI: 10.1142/S0218301317500781}
\begin{document}
\begin{frontmatter}
\markboth{\'A. Baran, T. Vertse}
{Matching polynomial tails to the cut-off Woods-Saxon potential}

\title{Matching polynomial tails to the cut-off Woods-Saxon potential}

\author{\'A. Baran}

\address{Faculty of Informatics, University of Debrecen, \\
PO Box 12, Debrecen,  H--4010,  Hungary\\
baran.agnes@inf.unideb.hu}

\author{T. Vertse}

\address{MTA Institute for Nuclear Research\\
PO Box 51, Debrecen, H--4001,  Hungary\\
vertse@atomki.hu}

\begin{abstract}
Cutting off the tail of the Woods-Saxon and generalized Woods-Saxon potentials changes the distribution of
the poles of the $S$-matrix considerably. Here we modify the tail of the cut-off Woods-Saxon (CWS) and 
cut-off generalized Woods-Saxon (CGWS) potentials by attaching Hermite polynomial  tails to them beyond 
the cut. The tails reach the zero value more or less smoothly at the finite ranges of the potential.
Reflections of the resonant wave functions can take place at different distances.  The starting points 
of the pole trajectories
have been reproduced not only for the real values and the moduli of the
starting points but also for the imaginary parts. 
\end{abstract}

\end{frontmatter}

\section{Introduction}
Gamow shell model (GSM)\cite{Mi09} became a useful tool in analyzing drip line nuclei produced
in laboratories with radioactive beam facilities. A most recent analysis of this type is  that 
of the $^7$Be(p,$\gamma$)$^8$B and the $^7$Li(n,$\gamma$)$^8$Li reactions in Ref.~\cite{Fo15}.
The key elements of GSM are the Berggren-ensembles of single particle
states. The   Berggren-ensemble might include resonant and sometime anti-bound
states beside the bound states and the scattering states along a complex path $\cal L$ of 
the wave number $k$ plane.
The shape of the path $\cal L$ determines the set of pole states  the $S$-matrix to be included 
in the ensemble.  Therefore to know where the poles are located on the $k$-plane has crucial importance. 

The member states of the Berggren-ensemble satisfy the radial Schroe\-din\-ger
equation:

\begin{equation}
\label{radsch}
u^{\prime\prime}(r,k)+[k^2 -\frac{l(l+1)}{r^2}-v(r) ]u(r,k)=0~,
\end{equation}
where prime denotes the derivative with respect to the radial distance $r$. The non-negative integer 
$l$ denotes the quantum number of the orbital angular momentum, $v(r)$ denotes the sum of the
nuclear and Coulomb potentials both having spherical symmetry.
In our case we have no Coulomb potential term.

\begin{figure}[th]
\centerline{\includegraphics[width=11.0cm]{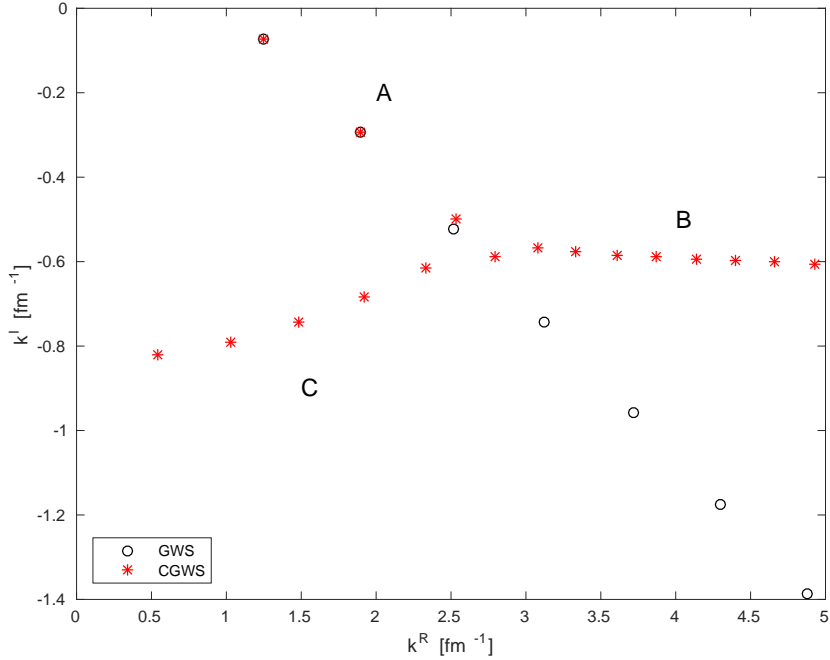}}
\caption{The poles of potential $GWS$ and $CGWS$ in the case of $R_{\max}=12$, 
$V_0=47.78$ MeV and $V_1=-200$ MeV.}
\label{abcgroup}
\end{figure}

The case of zero angular momentum ($l=0$) is very important, since in this case
the solution can be given in closed analytical form\cite{Be66} for generalized
Woods-Saxon (GWS) potential. The cut-off version of the GWS potential is denoted by CGWS potential, 
and that of the Woods-Saxon (WS) potential 
is by CWS. 
Some information  about the pole distribution of the
resonances in CWS and CGWS potentials was given in Ref.~\cite{Sa16} in comparison
that of the WS and GWS potentials without cut.
 In that work it was observed that for the CGWS potential with high potential barrier most of 
the resonances form groups denoted by A, B and C, see Fig. \ref{abcgroup}.
The resonances in group A
are due to the reflection of the solution at the radius of the potential (where the barrier is located), while
the resonances in groups B and C are resulted by reflections at the cut-off radius (group B) or 
the double reflections at the nuclear and the cut-off radii (group C). 
In the CWS potential without barrier only broad resonances were found
and reflections were at the cut-off radius of the potential, while at WS the reflection 
happened only at the potential radius.
 
In the present work we want to study the distribution of the poles
if we modify the tail of the CWS and the CGWS potentials by attaching polynomial tails to 
them instead of cutting them sharply. In this paper we restrict ourselves to the
$l=0$ case only where we have analytical solution in closed form\cite{Sa16} for the WS and 
GWS potentials without cut-off.

For CWS and CGWS potentials 
the $S$-matrix element of the given partial wave $l$ can be calculated from the
matching the solution being regular in the origin (r=0) to the asymptotical  solution
at $R_{as}$ i.e. at or beyond $R_{max}$, where the nuclear potential vanishes. 
In the asymptotical region our radial equation in Eq.(\ref{radsch}) evolves to its asymptotic form without potential and
it describes free spherical waves. These free waves satisfy the Riccati--Hankel differential equation 
\begin{equation}
\label{asimpde}
u^{\prime\prime}(r,k)+[k^2 -\frac{l(l+1)}{r^2} ]u(r,k)=0~.
\end{equation}
For a WS or GWS potential without cut-off  these potentials
vanish only at infinite distance and the matching to the asymptotical solution
can be performed at infinity. 
For nonzero angular momentum we have no accurate analytical
form for the solution of Eq.(\ref{radsch}) for WS and GWS potentials.   We are forced to use numerical integration
methods for calculating   $u_l(r,k)$ in the region $r\in [0,R_{as}]$.
Therefore we have to cut the potentials at finite distance from the origin (at $R_{max}$) and use CWS or CGWS forms.

At $r=R_{as}$ the numerical solution should match to that of the asymptotic equation. 
We can calculate the value of the scattering matrix $S_l(k)$ from the logarithmic derivative at $r=R_{as}$. 
Positions of the resonances are the pole positions of $S_l(k)$. Calculation of the poles of GWS potential is described in Ref.~\cite{Sa16}.
Examples for finding poles of $S(k)$ are given in Ref.~\cite{Bar15} .

\section{Nuclear potentials}
There are plenty of phenomenological nuclear potential forms used, but in this
work we restrict ourselves to the most popular form, the so called  Woods-Saxon potential and to its generalized form and 
 we focus attention to the effect of the radial change
of the nuclear potential in the region where it becomes zero.


The generalized WS potential (GWS) is a combination of a Woods-Saxon (WS) potential term and a surface term with potential
 strengths  $V_0$ and $V_1$, respectively. The radial form of the WS term is 
\begin{equation}
\label{WSshape}
f^{\rm WS}(r,R,a)=~-~
\frac{1}{1+e^{\frac{r-R}{a}}}~,
\end{equation}
while the shape of the surface term is 
\begin{equation}
\label{SWSshape}
f^{\rm SWS}(r,R,a)=~-~
\frac{e^{\frac{r-R}{a}}}{(1+e^{\frac{r-R}{a}})^2}~.
\end{equation}
The geometrical parameters of these terms are the radius $R$ and diffuseness $a$.  Therefore the resulting GWS potential is the following:
\begin{equation}
\label{GWSshape}
V^{GWS}(r,R,a,V_0,V_1)=V_0f^{WS}(r,R,a)+V_1f^{SWS}(r,R,a)~.
\end{equation}
The GWS potential is used not very often in actual nuclear calculations.
The surface term of it can produce potential barrier, therefore it can simulate
to some extent centrifugal or Coulomb barriers in a very approximate level. 

With $V_1=0$ we get the  Woods-Saxon potential:
\begin{equation}
\label{WS}
V^{\rm WS}(r,R,a,V_0)=V_0f^{\rm WS}(r,R,a)~.
\end{equation}

The cut-off generalized Woods-Saxon (CGWS) potential is a variant of the GWS potential in Eq. (\ref{GWSshape}), which
is more suitable for numerical calculation because it becomes zero and remains zero at and beyond a 
finite $R_{max}$ distance. These potentials are called 
strictly finite range (SFR) potentials and their range is the
distance $R_{max}$.

The CWS potential has the following form:
\begin{equation}
\label{WSpot}
V^{CWS}(r,R,a,R_{max},V_0)=V_0f^{CWS}(r,R,a,R_{max})~,
\end{equation}
where the radial shape is
\begin{equation}
\label{vagottWS}
f^{CWS}(r,R,a,R_{max})=-\left\{
\begin{array}{rl}
\frac{1}{1+e^{\frac{r-R}{a}}}
&\textrm{, if } r~<~R_{max}\\
0~~~~&\textrm{, if } r~\geq~ R_{max}~.
\end{array}
\right.
\end{equation}
 The CWS potential has  an additional parameter $R_{max}$, the cut-off radius. Beyond that finite distance the CWS form vanishes, and it helps in solving the
radial equation by numerical integration methods.
However, it is hard to assign any physical meaning to  
this extra parameter. The same is valid for the CGWS potential. 

In  Ref.~\cite{Sa16} we examined the effect of the sharp cut-off on the pole distribution. We found that the poles formed several 
groups as we shown in Fig. \ref{abcgroup}.

In this paper we modify the tail of these potentials, in order to check whether the change in 
the pole distribution was caused by  the sharp cut-off, or simply by the fact that the potential becomes to zero at a finite 
distance.

\subsection{Analytical behavior of the SFR potentials}

The $l=0$ states in the case of a SFR potential
\begin{equation}
\label{newpot}
V(r)=V_0~\theta({\cal R}-r)[({\cal R}-r)^\sigma +\ldots]~
\end{equation}
are discussed by R. G. Newton in his book\cite{Ne82} [see Eq.~(12.98) 
on p.~361 there]. Here $\sigma>0$, $\theta(x)$ denotes the 
Heaviside step function, and the square bracket contains a truncated 
expansion in terms of ${\cal R}-r$. In Eq.~(12.102) on p. 362 Newton gives 
the real and imaginary parts of the starting point $k_n=k_n^R +{\rm i} k_n^I$
of the trajectory of the $n$th pole of the $S$-matrix as follows:
\begin{equation}
\label{rek}
k_n^R= \frac{n\pi}{\cal R}+O(1)~,
\end{equation}
and
\begin{equation}
\label{imk}
k_n^I= -\frac{\sigma+2}{2{\cal R}}\ln (n) -O(1)~.
\end{equation}
Note that the dimensionless parameter $\sigma$  describes
the smoothness of the tail at the range $R_{max}={\cal R}$. 
Here we restrict ourselves to the integer values of parameter $\sigma$, which 
means a polynomial potential in Eq. (\ref{newpot}). Therefore we modify the tail of the potential 
to a polynomial function in the vicinity of ${\cal R}$.

The starting point of the pole trajectory is in the fourth quadrant of the 
$k$-plane.
In a shallow potential well there is no bound state. 
If we increase the strength to the usual value, we produce a few bound state
levels and the index $n$ gives the number of radial nodes in their wave functions.

Regge pointed out\cite{Re58} that a relation similar to Eq.~(\ref{rek}) is 
valid for the moduli of the starting wave number values: 
\begin{equation}
\label{absk}
|k_n|= \frac{n\pi}{\cal R}+O(1)=a_1 n+O(1)~.
\end{equation}

 In Ref.~\cite{Sa16} it was found that the resonances fall into three groups: A,B,C as it is shown in Fig. \ref{abcgroup}. The resonances in group A 
were the physical resonances caused by the reflection of the radial wave function at the  radius $R$ of the potential, 
where we have a large barrier due to the surface term in the CGWS potential.
The resonances in group B were due to the reflections at the cut-off radius $R_{max}$. The resonances in the third group (C) were caused by 
the reflections of the wave function at two distances, at the radius $R$ of the CGWS and at the cut-off distance  $R_{max}$ as well. 
The characteristic distance  for group C was approximately the difference $R_{max}-R$.
A  characteristic distance $\cal{R}$ (range) can be estimated  from the relation:
\begin{equation}
\label{range}
{\cal{R}}=\frac{\pi}{a_1}~, 
\end{equation} 
where the quantity $a_1$ is the slope of the best fit first order polynomial
\begin{equation}
\label{line}
p(m)=a_0+a_1m~.
\end{equation}
Here $m$ is the sequence number of the real part of the  decaying resonance pole (in the fourth 
quadrant of the $k$-plane). 
The best fit first order polynomial minimizes the sum of the
squares of the differences:
\begin{equation}
\label{deltamod}
\Delta(a_0,a_1,m_s,m_u)=\sum_{m=m_s}^{m_u} [|k_{m}|-p(m)]^2 \to {\rm min}~.
\end{equation}
By choosing the values of the lower cut ($m_s$) and upper cut ($m_u$) we can
select the poles included in the sum. To perform fitting makes sense only for at least three points, since two points fixes a straight line uniquely. 
 Naturally, the accuracy of the estimated distance might be influenced by the
nonlinearity of the polynomial in (\ref{line}). Nonlinearities can be caused by the interaction of the close lying poles\cite{Bar15} and by the accuracy
of the calculated $k_m$ values.

\section{Modifying the tail of the potential}

Now we study the change of the distributions of the poles caused by the changing
the tail of the CGWS potentials used in the calculation.
We consider the CGWS potential with the cut-off radius $R_f$. We would like to supplement this potential 
with different tails in order to examine the effect of the change of smoothness of the combined potential on 
the distribution of the resonances.  
We shall modify our notation and denote by  $R_{max}$ the distance where the tail of the potential disappears. 
The smoothness of the potentials  at $R_f$ and $R_{\max}$ can be characterized by two integers: $\nu$ and
$\mu$.  The value of the integer $\nu$ describes the
smoothness at $R_f$. The potential is continuous at $R_f$ for $\nu=1,2,3$, the first derivative of the potential is continuous for  $\nu=2,3$, while  
the  value $\nu=3$ denotes that even the second derivative of the potential is continuous at $R_f$. A similar classification can be introduced at 
the range  $R_{\max}$ by the integer $\mu$. The tail of the potential vanishes at    $R_{\max}$  for all the three values, i.e. for $\mu=1,2,3$. 
For $\mu=2,3$  the first derivative of the potential also disappears 
 at  $R_{\max}$, and the superscript  $\mu=3$ denotes that  the second derivative
 of the potential also disappears at $R_{\max}$.  
 The $f^{\nu H\mu}$ notation for the potentials considered in our work are summarized in Table 1.

\begin{table}[pt]
\caption{The potentials with polynomial tails we considered. The superscripts $\nu$ and $\mu$ 
measure the smoothness at $R_f$ and at $R_{\max}$, respectively }
\begin{center}
{\begin{tabular}{@{}cc|ccc@{}}\hline
&&\multicolumn{3}{ c }{$\mu$}\\
&&1&2&3\\
\hline
\multirow{3}{*}{$\nu$} & 1 & $f^{1H1}$&$f^{1H2}$&$f^{1H3}$ \\
& 2 & $f^{2H1}$&$f^{2H2}$&$f^{2H3}$ \\
& 3 & $f^{3H1}$&$f^{3H2}$&$f^{3H3}$ \\ \hline
\end{tabular}}
\end{center}
\end{table}

 In the case of $f^{3H2}$ potential we take a fourth order polynomial tail 
to obtain a twice continuously differentiable potential in $r\in (0,R_{\max})$ which disappears 
together with its first derivative at $R_{\max}$ ($R_{\max}> R_f$).

\begin{equation*}
f^{3H2}(r):=\begin{cases}
f^{GWS}(r)&\mbox{if } 0\leq r\leq R_f,\\
f^{H2}(r)&\mbox{if } R_f< r\leq R_{\max},\\
0&\mbox{if  } R_{\max}< r,
\end{cases}
\end{equation*}
where $f^{GWS}$ denotes the radial form of the GWS potential in Eq. (\ref{GWSshape}),
 while $f^{H2}(r)$ is a fourth-order polynomial, uniquely 
determined  by the five fitting conditions mentioned above, namely 
\begin{equation}\label{Hterm}
f^{H2}(r)=\left[\alpha+\beta (r-R_f)+\gamma (r-R_f)^2\right] (r-R_{\max})^2,
\end{equation}
where
\begin{align*}
\alpha &=\frac{f^{GWS}(R_f)}{(R_{\max}-R_f)^2}\\
\beta &=2\frac{f^{GWS}(R_f)}{(R_{\max}-R_f)^3}+\frac{(f^{GWS})'(R_f)}{(R_{\max}-R_f)^2}\\
\gamma &=3\frac{f^{GWS}(R_f)}{(R_{\max}-R_f)^4}+2\frac{(f^{GWS})'(R_f)}{(R_{\max}-R_f)^3}+\frac{(f^{GWS})''(R_f)}{2(R_{\max}-R_f)^2}~.
\end{align*}

The potentials $f^{2H2}$ and  $f^{1H2}$ can be obtained from the form of the potential  $f^{3H2}$ in Eq. \eqref{Hterm} by 
choosing $\gamma =0$ and $\beta =\gamma =0$, respectively. 
Remember, we denoted by $f^{2H2}$ the continuously differentiable 
potential  which disappears at $R_{\max}$ together with its first 
derivative (in this case the supplemented tail is a third order polynomial), while $f^{1H2}$ is 
a continuous potential for which $f^{1H2}(R_{\max})=(f^{1H2})'(R_{\max})=0$ remains true (the supplemented 
tail is a second order polynomial). 

The CWS potential can be considered as a special case of the CGWS potential with $V_1=0$ (without 
surface term).

Let us consider the potential $f^{3H1}$ which is twice continuously differentiable at $R_f$ and disappears at 
$R_{\max}$. (However, 
its first derivative is not equal to 0 at this point.) 

\begin{equation*}
f^{3H1}(r):=\begin{cases}
f^{GWS}(r)&\mbox{if } 0\leq r\leq R_f,\\
f^{H1}(r)&\mbox{if } R_f< r\leq R_{\max}~.\\
0&\mbox{if  } R_{\max}< r,
\end{cases}
\end{equation*}
The polynomial tail  $f^{H1}$ is the following third order polynomial
\begin{equation}\label{H1term}
f^{H1}(r)=\left[\alpha +\beta (r-R_f)+\gamma (r-R_f)^2\right] (r-R_{\max}), 
\end{equation}
with 
\begin{align*}
\alpha &=-\frac{f^{GWS}(R_f)}{(R_{\max}-R_f)}\\
\beta &=-\frac{f^{GWS}(R_f)}{(R_{\max}-R_f)^2}-\frac{(f^{GWS})'(R_f)}{(R_{\max}-R_f)}\\
\gamma &=-\frac{f^{GWS}(R_f)}{(R_{\max}-R_f)^3}-\frac{(f^{GWS})'(R_f)}{(R_{\max}-R_f)^2}-\frac{(f^{GWS})''(R_f)}{2(R_{\max}-R_f)}~.
\end{align*}

The smoothness at $R_f$ can be spoiled by choosing $\gamma =0$ or by $\beta =\gamma =0$. These expressions describe $f^{2H1}$ and $f^{1H1}$ (where 
$f^{2H1}$ is a continuously differentiable, while $f^{1H1}$ is a continuous potential, and both belongs to $\mu =1$, i.e. they 
disappear at $R_{\max}$).

Hence, we have given the exact forms of the potentials in the first two columns of the table. The 
 potentials in the third column of the Table 1 are the following. Let us define $f^{3H3}$ by the formula 

\begin{equation*}
f^{3H3}(r):=\begin{cases}
f^{GWS}(r)&\mbox{if } 0\leq r\leq R_f,\\
f^{H3}(r)&\mbox{if } R_f< r\leq R_{\max},\\
0&\mbox{if  } R_{\max}< r~.
\end{cases}
\end{equation*}
The polynomial tail $f^{H3}$ is the following fifth order polynomial
\begin{equation}\label{H3term}
f^{H3}(r)=\left[\alpha +\beta (r-R_f)+\gamma (r-R_f)^2\right] (r-R_{\max})^3,
\end{equation}
with 
\begin{align*}
\alpha &=-\frac{f^{GWS}(R_f)}{(R_{\max}-R_f)^3}\\
\beta &=-3\frac{f^{GWS}(R_f)}{(R_{\max}-R_f)^4}-\frac{(f^{GWS})'(R_f)}{(R_{\max}-R_f)^3}\\
\gamma &=-6\frac{f^{GWS}(R_f)}{(R_{\max}-R_f)^5}-3\frac{(f^{GWS})'(R_f)}{(R_{\max}-R_f)^4}-\frac{(f^{GWS})''(R_f)}{2(R_{\max}-R_f)^3}~.
\end{align*}

This formula results in a twice continuously differentiable potential which disappears together with its 
first two derivatives at the point $R_{\max}$. 

Similarly to the previous two groups, we obtain the potentials $f^{2H3}$ and $f^{1H3}$ by 
setting the values  $\gamma =0$ and $\beta =\gamma =0$, respectively.

\section{Numerical results}

We considered the system of  $^{56}$Fe with potential parameters taken from Ref.~\cite{Sa16} and Ref.~\cite{Ba15}, 
namely  $V_0=47.78$ MeV, $a=0.6$ fm, $R=4.92$ fm.

We determined the positions of the poles which formed different groups. For each groups we fitted straight lines to 
the values $|k_n|$  and calculated the range ${\cal R}$ in \eqref{absk} from the slope of the first order polynomial. 
From the values of the ranges we estimated the distance or distances where reflection or 
reflections  of the radial wave function take place. 

First we consider the $V_1=0$ case, when
we attach  different 
polynomial tails  to the CWS potential at $R_f=8.5$ fm, and examine the effect of tails disappearing at 
 $R_{\max}=10$ differently. 
Since $V_1=0$ we do not have any potential barrier. Due to this
 we have no narrow resonances, and the reflection at $R$ is not very pronounced. The wave function can be 
reflected mainly at $R_f$ and $R_{\max}$ depending on the smoothness of the potential at these distances.

\begin{figure}[th]
\centerline{\includegraphics[width=11.0cm]{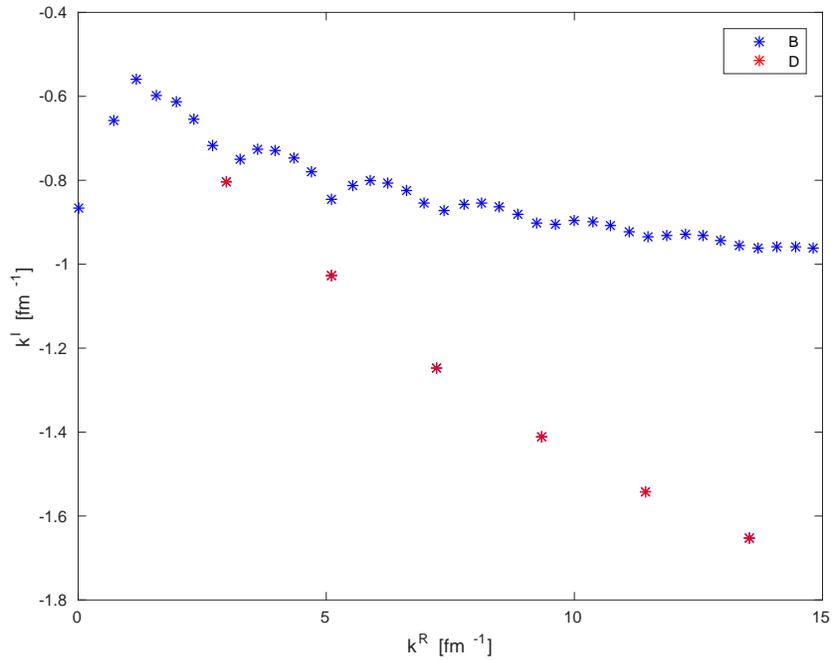}}
\caption{The poles of potential $f^{1H3}$ in the case of $R_f=8.5$, $R_{\max}=10$. 
Group D is separated from group B.}
\label{resG1H3}
\end{figure}

The potentials corresponding to $\nu =1$ are continuous, but not  
differentiable at $R_f$, i.e. the  first derivative jumps at $R_f$. Calculating the poles for different values of $R_f$ 
 and $R_{\max}$ we observe, that the resonant wave function can be reflected from this jump. Another reflection might happen at
the range of the potential, at $R_{\max}$. In Fig. \ref{resG1H3} we can see two
groups of resonances, the upper one is denoted by B, similarly to the case 
in  Ref.~\cite{Sa16}, while the new group is denoted by D (the red stars in the 
figure). 

\begin{figure}[th]
\centerline{\includegraphics[width=11.0cm]{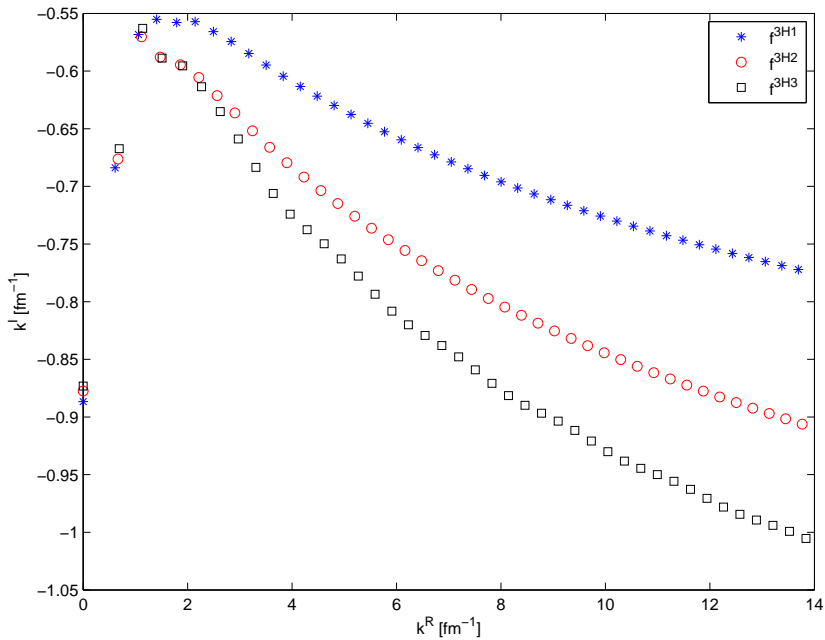}}
\caption{The poles of potential $f^{3H1}$, $f^{3H2}$, and $f^{3H3}$ in the case of $R_f=8.5$, $R_{\max}=10$.}
\label{res_nu3R10}
\end{figure}

 For $\nu =3$ the matching of the tail at $R_f$ is smooth enough  not to have reflections at $R_f$. 
In  Fig. \ref{res_nu3R10} the positions of the poles are shown for $\nu =3$ and for different $\mu$ values,   
the absence of the group D can be noticed in the figure.

However, at $R_{max}$ we cut the potential, so the wave function is reflected from $R_{max}$. 
Therefore the ranges are close to $R_{max}$=10 fm.

\begin{figure}[th]
\centerline{\includegraphics[width=11.0cm]{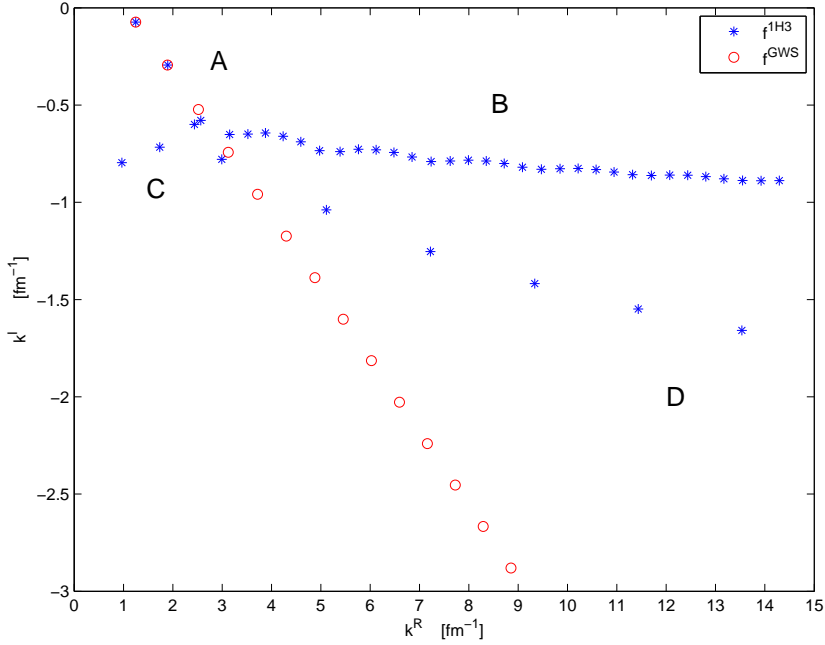}}
\caption{Pole positions for the potential $f^{1H3}$  in the case of $R_f=8.5$, $R_{\max}=10$, $V_0=47.78$, $V_1=-200$. 
The pole positions calculated for the GWS potential are also shown for comparison. }
\label{res1H3}
\end{figure}

\vfill\eject
\vfill\eject
\vfill\eject

Now we have a large barrier ($V_1=-200$) in which
narrow resonances might be formed. Reflections of the wave function can take place at the barrier at the radius $R$. 
These are explained in Ref.~\cite{Sa16} for the CGWS potential. The three groups of poles A, B and C (see Fig. \ref{abcgroup}) in that work are due to the 
reflections at $R$, $R_{\max}$ and $R_{\max}-R$, respectively. These three groups of poles are present also in the case when we attach a 
polynomial tail to the CGWS potential at distance $R_f$.
A fourth group (group D) also  appears when $\nu=1$ and $\mu=3$
holds as it does for $V_1=0$.
That means that in the case of $f^{1H3}$  four groups are present, the positions of the poles are shown in Fig. \ref{res1H3}. In order to compare 
to the poles of the analytical solution of the potential $f^{GWS}$ we plot the poles calculated in Ref.~\cite{Sa16}. The first two poles of the 
numerical solution in group A coincide with those of $f^{GWS}$, therefore they are not caused by the cut-off of the GWS potential. They are the true resonances in the GWS potential.
Two further poles in the GWS potential are close to that of the  $f^{1H3}$ potential, the last pole in group C and the first pole in group D.  
Anyway the further poles of the GWS potential are quite far from those of the potential $f^{1H3}$.
Therefore the combined effect of the cut-off and the modification of the tail are shown on the rest of the resonances in groups B,C,D.

\begin{figure}[th]
\centerline{\includegraphics[width=11.0cm]{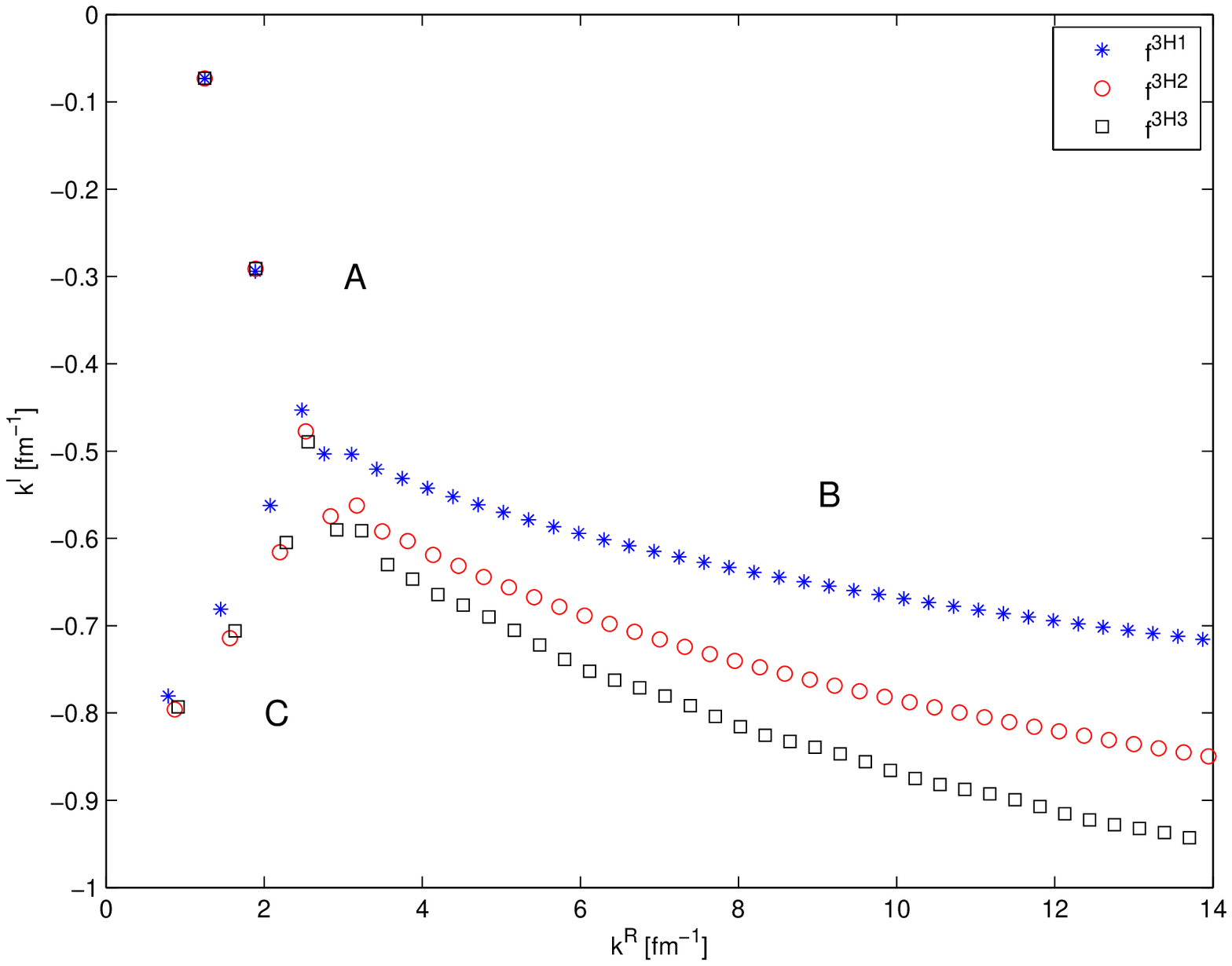}}
\caption{$\mu$-dependence of the pole positions for the potentials $f^{3H1}$, $f^{3H2}$   and  $f^{3H3}$ in the case of $R_f=8.5$ fm, $R_{\max}=10$ fm. 
The potentials strengths are  $V_0=47.78$ MeV and $V_1=-200$ MeV. }
\label{resnu3}
\end{figure}

In the case of $\nu =3$, when the tail is attached smoothly at $R_f$, and the group D is missing (see Fig.\ref{resnu3}), we  deduced ranges 
from Eq. \eqref{range} as we did before by fitting resonances in group A, B and C . The ranges obtained are summarized in Table 2 for different values of $R_{f}$  and $R_{\max}$. 
In Ref.~\cite{Sa16} we observed that 
ranges ${\cal R}_A$, ${\cal R}_B$ and ${\cal R}_C$ were equal approximately to $R$, $R_{\max}$ and $R_{\max}-R$, respectively. 
The values in Table 2 confirm the validity of this rule, even when we attache a polynomial tail to the GWS 
potential. In order to show it more explicitly we added a column with the values ${\cal R}_A+{\cal R}_C$. The values in this column 
are around the  value $R_{\max}$ in the last column.  
The deviations measure the uncertainty of our procedure. The deviation of the
deduced ${\cal R}_A+{\cal R}_C$ values from $R_{max}$ were always smaller than
1 fm.

\begin{figure}[th]
\centerline{\includegraphics[width=9.0cm]{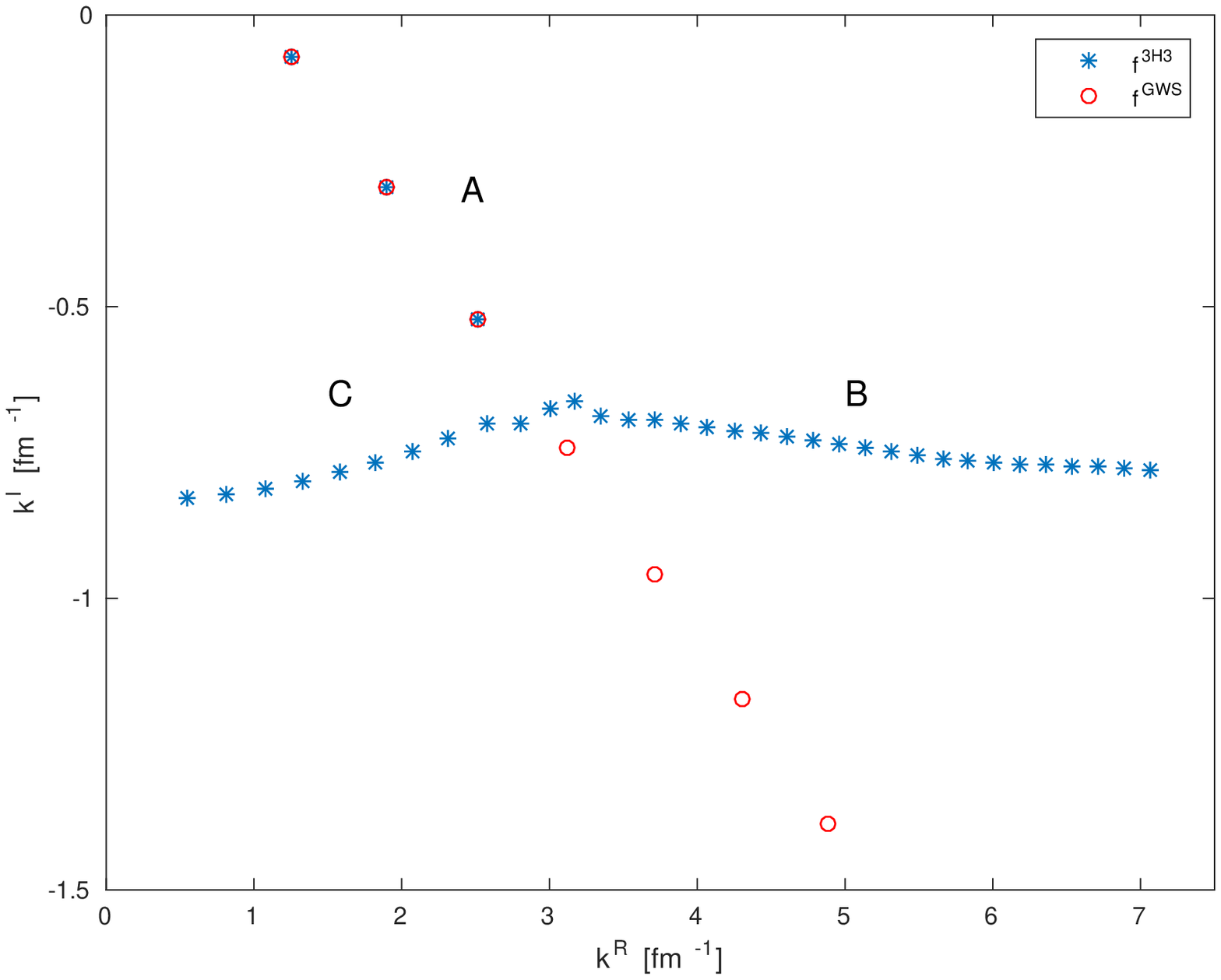}}
\caption{Pole positions for the potentials $f^{3H3}$ in the case of $R_f=17$ fm, $R_{\max}=18$ fm. 
The potentials strengths are  $V_0=47.78$ MeV and $V_1=-200$ MeV. }
\label{resnu3mu3}
\end{figure}

\begin{table}[pt]
\caption{Ranges  for $V_0=47.78$ MeV, $V_1=-200$ MeV . }
\begin{center}
{\begin{tabular}{@{}ccccccc@{}}\hline
Potential&${\cal R}_A$&${\cal R}_C$&${\cal R}_B$&$R_f$&${\cal R}_A+{\cal R}_C$&$R_{\max}$\\ 
\hline 
$f^{1H1}$&4.45&8.88&13.10&12.0&13.33&13\\
$f^{3H1}$&4.58&9.02&12.99&12.0&13.60&13\\
$f^{1H3}$&4.87&8.44&11.97&12.0&13.31&13\\
$f^{1H3}$&4.79&10.51&13.96&14.0&15.30&15\\
$f^{1H3}$&4.79&12.22&15.92&16.0&17.01&17\\
$f^{1H3}$&4.79&12.58&17.10&17.0&17.37&18\\
$f^{2H3}$&4.79&12.90&17.17&17.0&17.69&18\\
$f^{3H3}$&4.79&13.11&17.86&17.0&17.90&18\\
\hline
\end{tabular}}
\end{center}
\end{table}

If we use a higher value for $R_{max}$ and we keep the length of the tail fixed
for  $R_{max}-R_f=1.0$ fm, then the bifurcation of the poles to groups B and C
starts later and there are more poles in group C. The fourth pole in group A
is close to that of the pole in GWS potential, and the sum of the ranges
 ${\cal R}_A+{\cal R}_C$ is closer to the $R_{max}$ value. So the uncertainty
 of our method is reduced, see Fig. \ref{resnu3mu3}.

The new program JOZSO\cite{No17} we used here  made it possible to calculate the
positions of the resonances with higher accuracy, therefore we could test the imaginary parts $k^I_n$ of
the starting points of the pole trajectories given in Eq. (\ref{imk}).
In the work in Ref.~\cite{Sa14} a similar attempt failed because of the limited
accuracy of the program ANTI\cite{Ix95} used there.

We fitted a first order polynomial as a function of the $ln(n)$ values
\begin{equation}\label{line2}
p(n)=a_0+ a_1ln(n) 
\end{equation}
 
to the $|k^I_n|$ values, and used the relation:

\begin{equation}\label{sigma}
\sigma=2 R_{\max} a_1-2
\end{equation}
to calculate the values of the $\sigma$ parameter from the slope $a_1$ of the
polynomial in Eq. (\ref{line2}). 
We started counting of the poles from $m=4$ since the number of the
bound states is three in that potential well.

 \begin{figure}[ht]
\includegraphics[width=11.0cm]{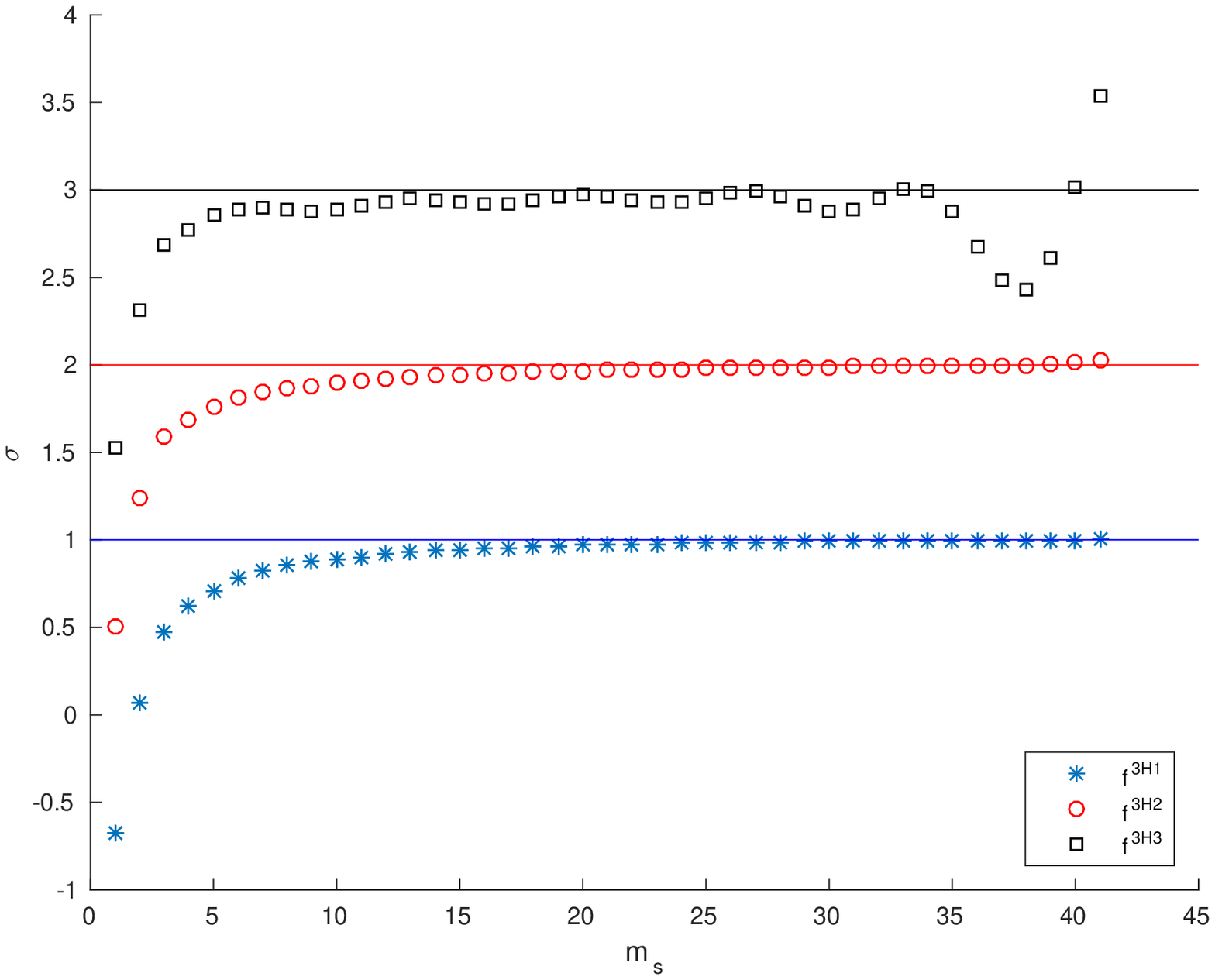}
\caption{$\sigma$ values for fitting the imaginary part of $f^{3H1}$, $f^{3H2}$ and $f^{3H3}$  in the case of $R_f=8.5$, $R_{\max}=10$. 
Potential strengths are $V_0=47.78$ MeV, $V_1=-200$ MeV.}
\label{sigmaV}
\end{figure}
 The results are displayed in Fig. \ref{sigmaV} for the potentials $f^{3H1}$, $f^{3H2}$ and $f^{3H3}$ with
strength $V_0=47.78$ MeV . The horizontal lines in the figure show the
$\sigma=\mu$ values.
The points approximate the corresponding $\sigma=\mu$ lines  as the lower cut value
$m_s$ increases reasonably well.
 These results confirm that
the precision of our new program (JOZSO) is superior to the programs we were
using before. The old program GAMOW\cite{Ve82} uses Fox-Goodwin method with
constant step length independently of the resonant state. The program ANTI\cite{Ix95} uses 
Ixaru's piecewise perturbation method\cite{Ix84}, which is a very efficient
numerical integration method in general, but its accuracy is controlled by the
change of the potential. It adjusts the step-length of the integration to the
prescribed change of the reference potential. In the program JOZSO 
an automatic adjustment of the integration step length for the individual resonance is more proper, 
and it describes better the very broad resonances concerned. This is reflected in the fact, that the 
convergence of the $|k^I|$ values for cases $\sigma=\mu>1$  can be well described by using the new program.

\section{Conclusions}
We have studied the distributions of the broad resonant poles in CWS and CGWS
potentials in which the tails of the potentials are modified by attaching
polynomial to the potentials.
We  can conclude that broad resonances appear as the result
of the reflection of the resonant wave function at the cut-off radius $R_{max}$.
The reflection is sensitive to the $\mu$ value, i.e. to the smoothness of the wave function at the $R_f$
distance, where we attach the polynomial tail to the original 
potentials.  If the joining of the tail is smooth $\nu=3$ then no reflection
at  $R_f$ can be observed. 
If the joining at  $R_f$ is not too smooth, i.e. $\nu=1$ or  $\nu=2$, then we can observe
the appearance an extra group of poles (group D) which can be attributed to the
reflection of the resonant wave function at $R_f$. 
Naturally we want a potential which is smooth everywhere within its range, therefore we should use $\nu=3$. 

If we have a large potential
barrier in the CGWS potential then the wave function can be reflected also from the
barrier resulting a few narrow resonances. These resonances form the group A
and the radius of the potential $R$ can be derived from the range $\cal{R}_A$
deduced from the slope of the best fit linear polynomial to the positions of the
poles in this group. This range   $\cal{R}_A$ does not depend on  $R_f$. Since 
the polynomial tail of the potential is unphysical (like the sharp cut-off of the CWS potential), we try to
limit its range to a small interval. At the outer end of the tail its value reaches zero depending on the parameter $\mu$. 
With increasing   $\mu$ the values of the first or the second derivatives reach zero continuously. Reflection at the $R_{max}$
distance, where the tail becomes zero can be deduced from the
distribution of the resonant poles with certain accuracy. The accuracy depends on the
errors of the numerical solution of the differential equation and also on the
assumption used to estimate the range. In our calculations these errors were
always below the 1 fm distance.

One has to appreciate that the new JOZSO program\cite{No17} is able to describe
the asymptotical values of the $k^I$ values given in Eq. (\ref{imk}).

\vfill\eject

\section*{Acknowledgements} 
This work was supported by the Hungarian Scientific Research Fund OTKA,
grant No. K 112962.


\begin{thebibliography}{10}
\bibitem{Mi09}
N. Michel, W. Nazarewicz, M. Ploszajczak and T. Vertse, {\it J. Phys. G.}  {\bf 36}, 013101    (2009).
\bibitem{Fo15}
K. Fossez, N. Michel, M. Ploszajczak, Y. Jaganathen, and R. M. Id Betan,  {\it Phys. Rev. C}  {\bf 91}, 034609 (2015).
\bibitem{Sa16}
P. Salamon, \'A. Baran, T. Vertse,
Nucl. Phys.  A, {\bf 952}, 1, (2016).
\bibitem{Ba15}
O. Bayrak and E. Aciksoz, {\it Phys. Scr.} {\bf 90}, 015302 (2015).
\bibitem{Be66}
Gy. Bencze, 
{\it Commentationes Physico-Mathematicae} {\bf 31}, 4 (1966).

\bibitem{Ne82} R.  G. Newton, \emph{Scattering Theory of Waves and Particles} (Springer, New York, 1982).

\bibitem{Sa14} P. Salamon, R. G. Lovas, R. M. Id Betan, T. Vertse, and L. Balkay,
{\it Phys. Rev. C} {89}, 054609 (2014).

\bibitem{Bar15}
\'A. Baran, Cs. Nosz\'aly, P. Salamon and T. Vertse,{\it European Physical Journal A}, {\bf 51}: 76 (2015).
\bibitem{Ve82}
T. Vertse, K. F. P\'al, Z. Balogh, {\it Computer Physics Communications} {\bf 27}, 309    (1982).

\bibitem{Ix95}
L.Gr. Ixaru, M. Rizea, T. Vertse,  {\it Computer Physics Communications} {\bf 85}, 217    (1995).

\bibitem{Ix84} L.Gr. Ixaru,  \emph{Numerical Methods for Differential Equations}, D. Reidel Publ. Comp. Dordrecht, Boston, Lancaster, 1984.

\bibitem{Re58}
T. Regge, {\it Il Nuovo Cim.} {\bf 8}, 671 (1958).


\bibitem{No17}
\'A. Baran, Cs. Nosz\'aly,  and T. Vertse, {\it JOZSO, a renewed version of the computer 
code GAMOW for calculating broad neutron resonances in phenomenological
nuclear potentials}, to be submitted to Computer Physics Communications.
\end{thebibliography}
\end{document}